\documentclass[showpacs,jcp,superscriptaddress,twocolumn]{revtex4-1}
\usepackage{amssymb}
\usepackage{amsmath}
\usepackage{mathrsfs}
\usepackage{stmaryrd}
\usepackage[dvips]{graphicx}
\usepackage{subfigure}
\usepackage[dvips]{color}
\newcommand{\mb}{\mathbf}
\newcommand{\mc}{\mathcal}

\newcommand{\ud}{{\rm d}}
\newcommand{\ue}{{\rm e}}

\begin{document}
\title{A hybrid particle-continuum resolution method and 
  its application to a homopolymer solution}

\author{Shuanhu Qi}
\affiliation{Institut f\"{u}r Physik, 
Johannes Gutenberg-Universit\"{a}t Mainz, D-55099 Mainz, Germany}
\affiliation{KITPC, Zhong Guan Cun East Street 55, P. O. Box 2735, Beijing 100190, P. R. China}

\author{Hans Behringer}
\affiliation{Institut f\"{u}r Physik, 
Johannes Gutenberg-Universit\"{a}t Mainz, D-55099 Mainz, Germany}
\affiliation{CUI, Universit\"{a}t Hamburg, 
Luruper Chaussee 149, D-22761 Hamburg, Germany}

\author{Thorsten Raasch}
\affiliation{Institut f\"{u}r Mathematik, 
Johannes Gutenberg-Universit\"{a}t Mainz, D-55099 Mainz, Germany}

\author{Friederike Schmid}
\affiliation{Institut f\"{u}r Physik, 
Johannes Gutenberg-Universit\"{a}t Mainz, D-55099 Mainz, Germany}

\begin{abstract}
We discuss in detail a recently proposed hybrid particle-continuum scheme for
complex fluids and evaluate it at the example of a confined homopolymer
solution in slit geometry.  The hybrid scheme treats polymer chains near the
impenetrable walls as particles keeping the configuration details, and chains
in the bulk region as continuous density fields.  Polymers can switch
resolutions on the fly, controlled by an inhomogeneous tuning function. By
properly choosing the tuning function, the representation of the
system can be adjusted to reach an optimal balance between physical accuracy 
and computational efficiency. The hybrid simulation reproduces the results
of a reference particle simulation and is significantly faster (about
a factor of 3.5 in our application example).
\end{abstract}

\maketitle

\section{Introduction}

In materials science, one often encounters situations where the relevant
properties of a material are crucially influenced by the local microstructure
at certain localized regions in space, whereas almost everywhere else, in the
``bulk'', the specific structure only matters in an average sense. For example,
details of the local structure at surfaces are more important for many
applications than details of the bulk structure \cite{polymer_surface}. Other
prominent examples are nanocomposites, where the behavior of a material can
be improved or even completely changed by filling it with small
quantities of additives \cite{nc1,nc2,nc3,protein-particle,nc4,nc5}. For a
physical modelling of such composites, a finer description of the additives,
e.g., at the atomistic level or even quantum level, is necessary, while a
coarser grained description of the remaining medium, e.g., at a  coarse-grained
particle level or continuum level, is often sufficient. This clearly
calls for multiscale hybrid simulation approaches.

Multiscale hybrid modelling is a challenging and rapidly developing field in
the communities of condensed matter physics, materials science, and engineering
\cite{QMMM1,QMMM2,atom-coarse,EEL07,Multi_FT,AA_CG,lockerby}. One prominent
early example is the quantum mechanics (QM)/molecular dynamics (MD) scheme
\cite{QMMM2}, where small (e.g., chemically reacting) regions are treated
at a quantum mechanical level and the environment by classical molecular
dynamics. Whereas particles cannot move from one resolution region to another
in the QM/MM scheme, the recently developed adaptive resolution schemes (AdResS)
\cite{review, adaptive} allow for free diffusion of particles and connect
different resolution regions in space by means of smooth interpolation
functions \cite{adaptive,potential,Lagrange}. AdResS schemes have been 
proposed that combine the hybrid atomistic/coarse grained scales \cite{AandC} 
and the quantum/coarse grained scales \cite{CandQ,Poma}. However,
these methods are restricted to particle-based simulation models.

On large scales, continuum descriptions which disregard microscopic details
(e.g., elastic models, phase field models, hydrodynamic models) are often
favorable, since they work with collective variables that directly reflect the
relevant mesoscopic properties of a material. Hence hybrid multiscale schemes
that couple small-scale particle-based descriptions to large-scale continuous
descriptions are of particular interest. One relatively straightforward
approach is to permanently treat certain components of a system as particles
immersed in a continuous medium \cite{IB_1,IB_2,BD_FTS,DPD_PF}. Another
increasingly popular type of approach is concurrent particle/field modeling,
where particle simulations are used to determine the dynamical evolution of a
continuum model.  Prominent examples are the heterogeneous multiscale method
\cite{EEL07,lockerby,HMM_01}, which uses particle simulations to adjust the
rheological parameters of a hydrodynamic continuum model on the fly, and
``Single chain in Mean Field'' simulation methods
\cite{BD_DSCF,SCF_BD_A1,SCMF1,SCMF2,SCF_MD1,SCF_MD2,SCF_MD3}, where simulations
of instantaneously uncorrelated chains are used to determine the dynamical evolution of fields in
a dynamic density functional. However, few methods exist that allow to ``zoom
into'' certain particle-resolved regions in space within a continuum simulation
in an adaptive resolution sense. One such scheme was developed by
Delgado-Buscalioni, de Fabritiis, and coworkers
\cite{particle_NS,particle_FH,MD_FHM,triple}. It can be applied to simple or
molecular fluids and couples molecular dynamics simulations of fluids with the
fluctuating Navier-Stokes equations \cite{fluctuation_hydrodynamics} through a
hybrid interfacial region. Another scheme which allows one to construct
adaptive resolution models for soft materials in a formally exact manner was
recently proposed by us \cite{Hybrid_PF}. The purpose of the present paper is
to discuss possible implementations and variants of this approach and to
evaluate it systematically at the example of a simple model system, a confined
homopolymer solution.

The construction of our hybrid scheme starts from particle resolution (PR)
models of the Edwards type \cite{Edwards}, in which the interactions are
expressed in terms of local densities. Edwards type models are popular starting
points in theoretical polymer physics
\cite{BD_DSCF2,RPA,GaussianApprox,variational,renormalization} and are often used for
the interpretation of experimental phenomena in soft matter and biophysics,
as well as for efficient particle-based computer simulations
\cite{Edwards01,Edwards02,Edwards03,Hans,brush_switch}. In our adaptive
resolution approach, we exploit the fact that the partition function of Edwards
models can be rewritten exactly as a fluctuating field theory. Hence each
Edwards particle model has a continuum partner, and our scheme relies on the
equivalence between the particle model and its continuum partner. This allows
one to single out selected regions in space to be treated at a particle level
within a field-based continuum simulation. 

We note that similar field theoretical mappings to continuous models can also
be performed for other particle models, as long as particles do not interact
with hard core interactions. Our scheme is not limited to Edwards models.
However, the underlying particle must be coarse-grained in the sense that
particle interactions are soft (or of Coulomb type). In the present paper, we
will specifically consider the simplest Edwards type model, which
describes homopolymer solutions in an implicit solvent. This model already
contains basic ingredients of soft matter systems, i.e., the competition of
entropy and interaction energy and the tunable and relatively large length
scales which characterize soft systems. 

We construct the hybrid adaptive resolution scheme following the basic idea
sketched in Ref.\ \cite{Hybrid_PF}, focussing on an implicit solvent system. We
explain in detail the construction and implementation of the hybrid scheme, and
evaluate the computational efficiency of the present scheme by comparison with
its pure particle-based counterpart. We hope that the present work will help
other researchers to implement the hybrid scheme.  The paper is organized as
follows. In Sec.\ 2, the hybrid model is derived, and numerical algorithms that
can be applied in practical simulations are discussed. Results for the confined
homopolymer solution are presented in Sec.\ 3. We summarize and conclude in
Sec.\ 4.

\section{Hybrid particle-continuum resolution model}

In this section, we present the basic methodology of the hybrid particle-field
simulation approach. For simplicity, we consider the simple system of a
homopolymer solution in an implicit solvent. Generalizations to other soft matter
systems are straightforward.

\subsection{Construction of the particle-continuum 
adaptive resolution method}\label{II_1}

Let us consider a collection of $n_t$ polymers in a three-dimensional volume
$V$. The polymers are modeled as linear chains of $N$ beads connected by
Gaussian springs \cite{GausianChain} with a spring constant $\frac{3}{b^2}$,
where we set $k_BT\equiv 1$ ($k_B$ is  the Boltzmann constant, $T$ the
temperature) and $b^2$ is the mean-squared bond length.  Throughout this paper,
we will measure lengths in units of the radius of the gyration of the ideal
chain, $R_g = \sqrt{\frac{Nb^2}{6}}$, hence $b^2 = \frac{6}{N}$.  The energy of
the system is expressed as an Edwards Hamiltonian \cite{offlattice}:
\begin{equation}
\label{eq:edwardsHamiltonian}
  \mc H =
     \frac{V}{n_t}\frac{N}{4}\sum_{\alpha=1}^{n_t}\sum_{j=1}^{N-1}
       \Big[\mb R_\alpha^j -\mb R_\alpha^{j-1}\Big]^2
     + \frac{v_0}{2}\int d\mb r\hat\phi^2_t(\mb r),
\end{equation}
where $\mb R_\alpha^j$ is the position of the $j$-th bead on the $\alpha$-th
polymer. The quantity $\hat\phi_t(\mb r)$ is the normalized density defined as
the microscopic bead density profile $\hat\rho_t(\mb r) =
\sum_{\alpha=1}^{n_t}\sum_{j=0}^{N-1}\delta(\mb r-\mb R_\alpha^j)$ divided by
the average bead density $\rho_0 = \frac{n_tN}{V}$. Since the density is
defined from the bead positions, the fundamental particle degrees of freedom
are fully resolved in (\ref{eq:edwardsHamiltonian}). The first term in the
Hamiltonian (\ref{eq:edwardsHamiltonian}) describes the probability
distribution of a Gaussian chain. The second term accounts for the effective
interaction between beads, and $v_0>0$ is the excluded volume parameter.  The
partition function in the canonical ensemble is given by
\begin{equation}\label{eq:Z1}
\mc Z=\frac{1}{n_t!\lambda_T^{3n_t}}
  \prod_{\alpha=1}^{n_t}\prod_{j=0}^{N-1} \int \ud \mb R_\alpha^j 
  \exp\Big[-\frac{n_t}{V}\mc H(\{\mb R\})\Big],
\end{equation}
where $\lambda_T$ is the thermal wavelength accounting for the
contribution from kinetic energy.  Physical quantities at equilibrium
 can be extracted from the partition function (\ref{eq:Z1}), for
instance by Monte Carlo methods \cite{Frenkel}. It can also serve as
the starting point for pure field theoretical simulations.

In the following we recast (\ref{eq:Z1}) into an equivalent hybrid
particle-field model in a mathematically exact way. The resulting model can
then be treated numerically, thus combining the advantages of particle based
Monte Carlo techniques and field theoretical simulations. This will be achieved
in three steps: (i) The chains will be divided into two virtual species in such
a way that the physics is unchanged. (ii) The two species will then be treated
by different representations, i.e., one species is still described by beads
whereas the second species will be described by fields. (iii) The field
representations will then be simplified by adopting approximations that
make a numerical treatment more feasible.

In the first step the identical chains of the system are partitioned into two
species which we call p-chain and f-chain. This can be done in several ways.
The strategy we adopt in the following introduces additional ``spin'' variables
$\tau_{\alpha j}$ for all the beads, each of which takes the values 1 or 0, and
a spatially varying conjugate ``potential'' $\Delta \mu({\bf r})$ which couples
to the spin of a bead at position ${\bf r}$.  The $\alpha$-th chain is defined
to be an f-chain if all $\tau_{\alpha j}$ are 0, otherwise it is a p-chain.
The potentials $\Delta \mu$ thus lead to a partitioning into $n_p$ and $n_f$
chains of p- and f-type.  Hereafter, we refer to $\Delta\mu$ as the tuning
function (TF).  The introduction of the partitioning and of the TF must not
change the physics of the system.  This is technically achieved by 
introducing the identities
\begin{equation}
\label{eq:idtau}
\sum_{\tau_{\alpha j}=0}^1 
  \ue^{\big[\Delta\mu(\mb R_\alpha^j)\tau_{\alpha j}
     -\ln(e^{\Delta\mu(\mb R_\alpha^j)}+1)\big]}=1,
\end{equation}
which are valid for each bead $j$ on each chain $\alpha$.
Inserting these identities into the partition function
(\ref{eq:Z1}), we get 
\begin{eqnarray}\label{eq:Z3}
  \mc Z&=&\frac{1}{n_t!\lambda_T^{3n_t}}
     \int \prod_{\alpha=1}^{n_t}\prod_{j=0}^{N-1} 
          \sum_{\tau_{\alpha j}=0}^1 \ud \mb R_\alpha^j  \:
   \nonumber\\
  && \times \:
     \ue^{\big[\sum_{\beta k}\big(\Delta\mu(\mb R_\beta^k)\tau_{\beta k}
          -\ln(e^{\Delta\mu(\mb R_\beta^k)}+1)\big)
   -\mc H(\{\mb R\}) \big]}.
\end{eqnarray}
By construction, this partition function is equivalent to the original
partition function (\ref{eq:Z1}), but the artificial variables $\tau_{\alpha
j}$ introduce two different types of chains, and an inhomogeneous TF will
generate an inhomogeneous partitioning in space. This, however, does not change
the physics of the system. By adjusting the form of $\Delta\mu(\mb r)$, one can
manipulate the individual density distribution for p-chains and f-chains.  For
example, larger values of $\Delta\mu(\mb r)$ in some regions will result in
higher densities of p-chains in these regions, while smaller values of
$\Delta\mu(\mb r)$ will result in higher density of f-chains. By construction
the sums of the densities for p-chains and f-chains are the same independent of
the choice of $\Delta\mu(\mb r)$ and are identical to the ones obtained
directly from (\ref{eq:Z1}).

In the partitioning scheme discussed above, $p$-chains and $f$-chains have a
very different entropy in spin space: A p-chain can support $2^N-1$ possible
combinations of spin variables, while only one spin combination is possible in
f-chains. To compensate for this asymmetry and achieve $\langle n_p\rangle \sim
\langle n_f \rangle$, the values of the TF must be of order
$\Delta\mu_0=\ln[2^{1/N}-1]$.  Alternative ways to assign virtual identities
that avoid this asymmetry are also conceivable. For example, one could attach a
spin variable $\tau_{\alpha 0} \in \{0,1\}$ only to the first bead of the
$\alpha$-th chain, and declare this chain to be a p- or f-chain if
$\tau_{\alpha 0}=1$ or $0$, respectively. In this scheme, the identity
(\ref{eq:idtau}) is only inserted for beads $j=0$ on chains $\alpha$.  The
space-dependent tuning function $\Delta \mu $ still determines the spatial
distribution of the p- and f-chains and their total numbers $\langle n_p
\rangle$ and $\langle n_f \rangle$, but now the partitioning into p-chains and
f-chains is symmetric: The local ratio of f- and p-chains can be exactly
reversed by reversing the sign of $\Delta \mu(\mb r)$ and at $\Delta\mu(\mb
r)\equiv 0$, the p- and f-chains have the same density distribution. Despite
the advantages of such a symmetric scheme, we will use the asymmetric version
described earlier, because it treats all beads on the chain on equal footing.

The above construction partitions all chains into either p-chains or f-chains,
but they are both still described by particle degrees of freedom.  However, we
are free to choose other representation methods. One possibility is to describe
all the chains by continuum field models. The transition from PR models to field resolution (FR) models can be done in a
formally exact way using field theoretic methods (see below), and
field-theoretic simulation methods have proved to be powerful tools to
investigate the properties of polymeric systems\cite{FTS,Fredrickson}. In the
hybrid model, we describe only the f- chains by the FR method and the
others still by the PR method. The advantage of such an scheme lies in
its ability to exploit and combine the specialties and advantages of different
representation schemes simultaneously. 

To this end, we first reorder the sum over chains $\alpha$ such that the
$p$-chains come first.  The conversion of f-chains to a field representation is
then done following a standard field-theoretical method, i.e., the fields are
introduced through inserting an identity into the partition function
\cite{Schmid,SCMF}
\begin{equation}
1\propto\int\mc D W_f\mc D\phi_f
  \exp\left\{\frac{n_t}{V}\int d\mb r \: i W_f(\mb r)
  \Big[\phi_f(\mb r)-\hat\phi_f(\mb r)\Big]\right\},
\end{equation}
where $\hat\phi_f(\mb r) = \frac{V}{n_t N}
\sum_{\alpha=n_p+1}^{n_t}\sum_{j=0}^{N-1}\delta(\mb r-\mb R_\alpha^j)$
is the configurational dependent density of all beads belonging to f-chains, 
$\phi_f(\mb r)$ is the fluctuating density field corresponding to
these f-chains, and $W_f(\mb r)$ is the fluctuating auxiliary
field conjugate to the density field. Inserting this identity effectively
decouples the particle degrees of freedom of f-chains such that they can
be integrated out. Each f-chain then contributes the same factor 
$Z_0 \: Q_f$ to the partition function, where $Q_f$ is the 
normalized single-chain partition function 
\begin{eqnarray}\label{eq:Qf}
\lefteqn{Q_f[i W_f]= \frac{1}{Z_0} \int d\mb R_0\cdots d\mb R_{N-1}} \quad &
\nonumber\\
&\times \exp\Big\{ &-\frac{N}{4}
  \sum\limits_{j=1}^{N-1}(\mb R_j-\mb R_{j-1})^2-\frac{1}{N}
  \sum\limits_{j=0}^{N-1}i W_f[\mb R_j]\nonumber\\
&&- \:\sum\limits_{j=0}^{N-1}\ln[e^{\Delta\mu(\mb R_j)}+1]\Big\},
\end{eqnarray}
(the subscripts $0, \cdots, N-1$ refer to the indices of beads within a chain).
The normalization constant $Z_0 =  V\big(\frac{N}{4\pi}\big)^{3(N-1)/2}$ corresponds to
the partition function of an ideal noninteracting reference chain of length $N$
and was introduced to ensure that $Q_f$ remains finite in the limit of $N \to
\infty$ \cite{Fredrickson}.  The partition function (\ref{eq:Z3}) then reads
\begin{eqnarray}
\label{Z2}
\mc Z&=&\frac{1}{n_t!\lambda_T^{3n_t}}
  \sum_{\{\tau_{\alpha j} \}}
  \int \mc D W_f(\mb r)
  \int \mc D\phi_f(\mb r) 
  \prod_{\beta=1}^{n_p}\prod_{k=0}^{N-1} \int \ud \mb R_\beta^k
  \nonumber \\
 && \times \; \exp\Big\{-H_\tau -\frac{n_t}{V}\mc F\Big\},
\end{eqnarray}
where the sum $\sum_{\{\tau_{\alpha j}\}}$ runs over all possible
spin configurations ($\tau_{\alpha j} = 0,1$), the variable $n_p$
depends on the spin configuration as explained above,
the spin Hamiltonian $H_\tau$ is defined by
\begin{equation}
\label{eq:Htau}
H_\tau = 
- \sum_{\alpha=1}^{n_t}\sum_{j=0}^{N-1}
  \Delta\mu(\mb R_\alpha^j)\tau_{\alpha j}
+ \sum_{\alpha=1}^{n_p}\sum_{j=0}^{N-1}
   \ln\Big[e^{\Delta\mu(\mb R_\alpha^j)}+1\Big],
\end{equation}
and the effective free energy energy functional for polymers $\mc F$ 
is given by
\begin{eqnarray}
\label{eq:finalF}
\mc F [\{ \mb R_\alpha^j \}, \phi_f, i W_f]
  =  \mc F_p+\mc F_f+\mc F_\mathrm{mix}
\end{eqnarray}
with 
\begin{eqnarray*}
\mc F_p & = &
   \frac{V}{n_t}\frac{N}{4}\sum_{\alpha=1}^{n_p} \sum_{j=1}^{N-1}
    \Big[\mb R_\alpha^j-\mb R_\alpha^{j-1}\Big]^2
      +\frac{v_0}{2}\int d\mb r\hat\phi_p^2(\mb r)\\
\mc F_f & = &
   \frac{v_0}{2}\int d\mb r\phi_f^2(\mb r)
    - \int d\mb r \: i W_f(\mb r)\phi_f(\mb r)\\
   && - \: \frac{Vn_f}{n_t}\ln Q_f[i W_f] 
      - \: \frac{Vn_f}{n_t}\ln Z_0 \nonumber\\
\mc F_\mathrm{mix} &=& 
   v_0\int d\mb r\hat\phi_p(\mb r)\phi_f(\mb r)\nonumber.
\end{eqnarray*}
Here $\mc F_p$ is the Hamiltonian of the pure p-chains, $\mc F_f$ is the free
energy of the pure f-chains, and $\mc F_\mathrm{mix}$ is the coupling term.
The final expressions (\ref{eq:finalZ}) of the partition function and
(\ref{eq:finalF}) for the free energy of a configuration represent our hybrid
model which we will further investigate in this article. We note again that
there are two kinds of independent degrees of freedom left in the model. One
refers to the configurational space which belongs to the PR part, the other is
the auxiliary field which belongs to the FR part.

Now we work simultaneously with two kinds of representations, the PR and the FR
coexisting in the very same system. Within the PR, polymer chains are
characterized by the positions of beads, and the detailed information about
configurations is still accessible. Consequently the PR part is referred to as
a high resolution representation. The FR of f-chains corresponds to a lower
resolution since the configurational information on individual chains is lost,
only the continuous density can be observed. The regions in space where the
high and the low resolution is adopted are adjusted by the choice of $\Delta
\mu$ which governs where the respective representation prevails.

The combined particle-field approach is constructed such that it is formally
equivalent to the pure particle model, due to the fact that the field
description is introduced by an identity transformation/operator in the
partition function. Independent of the choice of the TF, one should obtain the
same density distributions for both the pure particle model and the hybrid
model.  In reality, however, the two models are not truly equivalent, since
simulations of the field model inevitably must make approximations. Most
importantly, the space must be discretized.  This amounts to a regularization
of the field theory which is necessary both for obvious practical reasons and
for fundamental reasons, to eliminate ultraviolet divergences in the theory
\cite{katz_05}. The FR model is thus necessarily coarse-grained with a
coarse-graining cutoff that is set by the grid size (which  gives us one more
justification to refer to the FR representation as the ``low resolution''
representation).  We note that the finite cutoff leads to a renormalization of
model parameters both in the field theory \cite{wang_02,kudlay_03,morse_07} and
in particle simulations with density-based Edwards-type interactions
\cite{brush_switch} (since the density dependent interactions are also
evaluated on a grid, see Sec.\ \ref{Simulation}). 

In practical applications of the hybrid model, further approximation
may be necessary at the FR level to achieve the desired speedup of the 
simulation time. They are discussed in the next section. 

\subsection{Treatment of the field degrees of freedom}\label{fieldPart}

The hybrid particle-field model has particle degrees of freedom (the bead
positions) and field degrees of freedom (density fields and auxiliary
potentials $W$). One difficulty arises from the fact that the free energy
functional of the fields $\mc F_f$ (Eq.\ (\ref{eq:finalF})) becomes complex.
Simulations of pure FR models are nevertheless possible with the Complex
Langevin technique \cite{ComplexLangevin,FTS}, however, they become cumbersome
and time consuming. In a hybrid simulation, the Complex Langevin simulation
scheme would involve a random walk of {\em all} degrees of freedom including
the particle degrees of freedom in the whole complex plane. This would
seriously affect the PR part of the simulation.  In order to avoid that
problem, we resort to integrating out some or all field degrees of freedom
beforehand within a saddle point approximation. This has the additional
advantage that it significantly speeds up the simulations. 

One possibility is to carry out the saddle point integration over $W_f(\mb r)$
and keep the explicit integral over the ''density'' field $\phi_f(\mb r)$.
Minimizing $\mc F [\{ \mb R_\alpha^j \}, \phi_f, i W_f]$ with respect to
$W_f(\mb r)$ gives the following relation
\begin{eqnarray}\label{eq:density}
\phi_f(\mb r)&\stackrel{!}{=}&
  -\frac{Vn_f}{n_tQ_f}\frac{\delta Q_f}{\delta i W_f(\mb r)}
   \Bigg|_{W_f^*}\nonumber\\
  &=&\frac{Vn_f}{n_t}
   \frac{\sum_{j=0}^{N-1} q_j(\mb r) \: q_{N-1-j}(\mb r)}
     {N\int \ud \mb r \: q_0(\mb r) \: q_{N-1}(\mb r)}
\end{eqnarray}
between $\phi_f(\mb r)$ and the saddle point solution $W_f^*(\mb r)$.
Here $Q_f$ is given by Eq.~(\ref{eq:Qf}) and $q_j(\mb r)$ is the end-integrated
propagator for a chain, which can be computed via the following recursion 
relation:
\begin{eqnarray}
q_0(\mb r)&=&  \ue^{-i \frac{1}{2} W_{\mbox{\scriptsize eff}}^*(\mb r)} \\
q_j(\mb r)&=& C 
\ue^{-\frac{1}{2} i W_{\mbox{\scriptsize eff}}^*(\mb r)}
\int \ud\mb r'\ue^{-\frac{N}{4}(\mb r-\mb r')^2}
 \ue^{-\frac{1}{2} iW_{\mbox{\scriptsize eff}}^*(\mb r')} 
  \:  q_{j-1}(\mb r') 
\nonumber \\
&& \quad \mbox{for $j>0$}
\end{eqnarray} 
with $C = \sqrt[3]{N/4 \pi}$ and $i W_{\mbox{\scriptsize eff}}^* := i
W_f^*+\ln[e^{\Delta\mu}+1]$. Obviously, one has $\int \ud\mb r\phi_f(\mb r)=Vn_f/n_t$.
Furthermore, the single chain partition function (\ref{eq:Qf}) can be expressed
in terms of the propagator by $Q_f=\frac{1}{V}\int \ud \mb r \: q_0(\mb r) \:
q_{N-1}(\mb r)$.  Since the end-integrated propagators are convolutions, they
can be calculated efficiently using fast Fourier transforms.  Note that the
propagator has been carefully constructed such that all beads, including the
end beads, carry the full weight $\ue^{-i W_\mathrm{eff}^*(\mb r)}$. This is
essential especially when considering very short chains.  In the continuum
approximation the end-integrated propagator satisfies the so called modified
diffusion equation \cite{Edwards,Freed}
\begin{equation}
\frac{\partial q(\mb r,s)}{\partial s}=\nabla^2 q(\mb r,s)
  -iW_{\mbox{\scriptsize eff}}^*(\mb r)q(\mb r,s),
\end{equation}
with $s \in [0:N]$, which is 
commonly used in field theoretical simulation models
(except for the term $\ln[e^{\Delta\mu}+1]$).

The saddle point integration with respect to the fields $W_f$ solves the
problem of the complex free energy functional: The saddle field $W_f^*$ is
purely imaginary, hence $i W_f^*(\mb r)$ is real and the saddle point
functional $\mc F^* [\{ \mb R_\alpha^j \}, \phi_f] =\mc F [\{ \mb R_\alpha^j
\}, \phi_f, i W_f^*] $ becomes real.  Nevertheless, the evaluation of
$\mc F^*$ for a given configuration of the field $\phi_f$ remains cumbersome,
since one has to solve the implicit equation Eq.\ (\ref{eq:density}) for $i
W_f^*$. To simplify the numerical calculations, we perform a variable transform
$\int \mc D \phi_f \to \int \mc D\omega_f$ in the partition function with
$\omega_f:= i W_f^*$. This variable transformation is associated with a
Jacobian, which can be incorporated in an additional contribution to $F$.
However, this contribution is of the same order of magnitude than the leading
(Gaussian) fluctuation correction to the saddle point integral, and will be
neglected as well. The final expression for the partition function is given by
\begin{eqnarray}
\label{eq:finalZ}
\mc Z&=&\frac{1}{n_t!\lambda_T^{3n_t}}
  \sum_{\{\tau_{\alpha j} \}}
  \int \mc D \omega_f(\mb r)
  \prod_{\beta=1}^{n_p}\prod_{k=0}^{N-1} \int \ud \mb R_\beta^k
  \nonumber \\
 && \times \: \exp\Big\{-H_\tau -\frac{n_t}{V}\mc F \Big\},
\end{eqnarray}
where $H_\tau$ and $\mc F = \mc F [\{ \mb R_\alpha^j \}, \phi_f, \omega_f]$ are
defined in Eqs.\ (\ref{eq:Htau}) and (\ref{eq:finalF}) and $\phi_f$ can be
calculated from $\omega_f$ according to Eq.\ (\ref{eq:density}) with $\omega_f
= i W_f$. This summarizes our hybrid model which we will further investigate in
this article. We note again that the model operates with two types of
independent degrees of freedom: The chain configurational space which belongs
to the PR part, and the auxiliary field space which belongs to the FR
part.

Alternatively, we can also carry out the saddle point integration with respect
to both fluctuating fields, $W_f$ and $\phi_f$. Hence $\mc F$ is minimized
with respect to both fields, which results in two saddle point equations,
Eq.\ (\ref{eq:density}) and
\begin{equation}
\label{eq:saddle_omega}
 i W_f^*(\mb r) = \omega_f^*(\mb r) 
 \stackrel{!}{=} v_0 \: \big( \phi_f^*(\mb r)+\hat\phi_p(\mb r)\big).
\end{equation}
Within this approximation, fluctuations in the FR part of the model
are fully neglected.

\subsection{Simulation method}\label{Simulation}

We use the Monte Carlo method \cite{Frenkel} to sample the remaining degrees of
freedom and to calculate statistical averages for the quantities of interest of
within the hybrid model. Suitable update moves therefore pertain to the
configuration $\{\mb R_\alpha^j\}$ of the PR part, the field $\omega_f(\mb
r)$ of the FR part, and the virtual spin configuration $\{ \tau_{\alpha j} \}$. 

It is important to note that we must introduce a spatial grid already at the
level of the particle model in order to evaluate the density dependent
interactions in Eq.\ (\ref{eq:edwardsHamiltonian}). The grid and the
prescription for calculating the local bead density $\hat{\phi}$ on the grid
from a (continuous) chain configuration are part of the definition of the
particle model. 

Here we split the physical space into small cubic cells whose centers define a
grid of points $r_{ijk}$, here, $i,j,k$ denote the index of a grid point. A
discretized density p-chain field is then defined on this grid through a
``particle-to-mesh" technique \cite{PM}. For simplicity, we use the
``nearest-grid-point" scheme, i.e., $\hat\phi_p(r_{ijk})$ is defined as the
average density in the cell. The definition of the Hamiltonian 
contains self-interactions. For the near-grid-point scheme, the self-energy is a
position independent constant \cite{self_energy}, and it only shifts the chemical potential. 
In the present case, this shift can be completely adsorbed in the tuning function 
and will not affect the statistical averages. The continuous TF is also discretized on the
grid: A bead in the cell with center point $r_{ijk}$ is subject to the TF
$\Delta\mu(r_{ijk})$. In order to have a consistent definition of collective
variables in the FR and PR domains, the field degrees of freedom $\omega(\mb
r)$ are discretized on the same grid.  However, this is not strictly necessary,
and more flexible schemes where the FR grid is much coarser than the
PR grid are conceivable.

A simulation starts with an initial configuration $\{\mb R_\alpha^j\}$ for
$n_p$ p-chains and an auxiliary potential $\omega_f(\mb r)$ describing the
$n_f$ f-chains. All p-chains are initially generated as Gaussian chains, and
the initial configuration of $\omega_f(\mb r)$ can be set randomly. The initial
number of $n_p$ and $n_f$ can be chosen at will, just respecting the total
number constraint $n_p + n_f = n_t$. Different starting values will not affect
the statistical quantities, but influence the equilibration time of the system.
Since the positions of beads and the potential are independent of each other,
their respective configurations are updated separately. A Monte Carlo move
comprises three steps: (i) The configuration of the p-chains is updated $M_p$
times, keeping the potential fixed; (ii) The auxiliary potential of the
f-chains is updated $M_f$ times, keeping the p-chain configuration fixed; (iii)
An identity switch between p- and f-chains is attempted $M_s$ times, keeping
the configuration of the other chains and the potential fixed.  All these trial
update steps are accepted according to the Metropolis rule.

\subsubsection*{i. p-chain update}

In updating the p-chain configurations, we randomly select one bead on a
randomly chosen p-chain, then move it by a random vector with a length
comparable to the bond length. For the end beads, one can also perform
reptation moves, meaning that end beads are removed and re-attached to the
other end, with a bond vector generated from a Gaussian distribution function
with variance given by the squared bond length. 

For beads $j$ on chains $\alpha$, trial displacements from the position 
$\mb R_\alpha^j$ to $\tilde{\mb R}_\alpha^j$ are accepted with the Metropolis 
probability
\begin{equation}
P_p=\mbox{min}\Bigg\{1, \exp\Big[- \Delta H_\tau
    -\frac{n_t}{V}(\Delta\mc F_p
    +\Delta\mc F_{\mbox{\scriptsize mix}})\Big]\Bigg\},
\end{equation}
where $\Delta H_\tau=-\Delta\mu(\tilde{\mb R}_\alpha^j)
  +\Delta\mu(\mb R_\alpha^j) 
  +\ln[e^{\Delta\mu(\tilde{\mb R}_\alpha^j)}+1]
  +\ln[e^{\Delta\mu(\mb R_\alpha^j)}+1]$, and
$\Delta\mc F_p=\tilde{\mc F}_p-\mc F_p$, $\Delta\mc F_{\mbox{\scriptsize
    mix}}=\tilde{\mc F}_{\mbox{\scriptsize mix}}-\mc
F_{\mbox{\scriptsize mix}}$ are the free energy differences associated
with this trial move. In case of the reptation move, $\Delta\mc F_p$
has to exclude the term associated with the bead connectivity (Gaussian
spring), since it is already taken into account by the generation of
the new bond vector.

\subsubsection*{ii. f-chain update.}

As discussed in Sec.\ \ref{fieldPart}, we perform partial or full saddle
point integrations with respect to the field degrees of freedom. This 
is implemented with the following sampling schemes:

\paragraph*{Partial saddle point approximation.}
In this case, the ``density'' degrees of freedom are still fluctuating
and our aim is to integrate the partition function (\ref{eq:finalZ}) by
sampling the auxiliary field variables $\omega_f$. A simple way to 
implement this in a Monte Carlo scheme is to generate trial field
values by adding a (small) random (positive or negative) number to the 
old value at each grid point $(ijk)$, e.g.,
\begin{equation}
\tilde\omega_f(r_{ijk})=\omega_f(r_{ijk})+h \cdot  a_{ijk}
\end{equation}
where the random numbers $a$ are generated uniformly and range
from -1 to 1. The magnitude of the constant $h$ affects the
equilibration time of the system. The acceptance probability for this
move is 
\begin{equation}
  P_f = \mbox{min}\Bigg\{1, \exp\Big[-\frac{n_t}{V}
  (\Delta\mc F_f+\Delta\mc F_{\mbox{\scriptsize mix}})\Big]\Bigg\}
\end{equation}
where $\Delta\mc F_f=\tilde{\mc F}_f-\mc F_f$ and 
$\Delta\mc F_{\mbox{\scriptsize mix}}$ are the free energy differences 
associated with this move. Below, we will refer to this potential 
unbiased algorithm as the PU scheme.

To improve the acceptance rate, one can adopt a potential biased scheme, where
the trial displacement vectors $ \{ a_{ijk} \}$ are biased towards the free
energy gradient $\delta F_f/\delta \omega_f$\cite{MC_bias}. Unfortunately, the
calculation of the gradient in our case is cumbersome: Using the relation
$\phi_f[\omega_f]$ defined by Eq.\ (\ref{eq:density}), we can rewrite the
gradient as
\begin{displaymath}
\frac{\delta F_f}{\delta \omega_f(\mb r)} = 
\int \: \ud \mb r' \: \frac{\delta F_f}{\delta \phi_f(\mb r')} \:
\frac{\delta \phi_f(\mb r')}{\delta \omega_f(\mb r)},
\end{displaymath} 
where $\delta\mc F/\delta\phi_f$ can be obtained easily {\em via}
\begin{equation}
\label{eq:phiforce}
\frac{\delta\mc F}{\delta\phi_f (\mb r) }
  =v_0 \big(\phi_f (\mb r)+\hat\phi_p(\mb r)\big)-\omega_f (\mb r)
  =: F(\mb r),
\end{equation}
but $\delta \phi_f (\mb r')/\delta \omega_f(\mb r)$ is a two-point correlation
function \cite{Schmid} whose exact calculation can be time consuming.
Therefore, we use a scheme that biases the trial moves of $\omega_f$
in the direction of $\delta\mc F/\delta\phi_f$. This scheme will be denoted
PB scheme. Specifically, the trial moves are constructed via
\begin{equation}
\tilde{\omega}_f = \omega_f + h \cdot \:F+\xi,
\end{equation}
where $F$ is given by Eq.\ (\ref{eq:phiforce}) and $\xi$ is a random
number satisfying $\langle\xi_n\rangle=0$ and 
$\langle\xi_n\xi_{n'}\rangle=2 h$. The priori
transition probability from $\omega_f$ to $\tilde{\omega}_f$ is
then given by
\begin{equation}
P^{\mathrm{prior}}_{\omega_f\to\omega'_f}
\propto \exp\Big[-\frac{\big(\tilde{\omega}_f-\omega_f- h \: F\big)^2}{4h}\Big],
\end{equation}
hence auxiliary potentials $\tilde{\omega}_f$ that approach the saddle point
solution are sampled with higher probability. To recover detailed balance,
the acceptance probability for the biased move is evaluated as
\begin{equation}
P_f = \min\left\{1, 
   \frac{P^{\mathrm{prior}}_{\tilde{\omega}_f\to\omega_f}}
        {P^{\mathrm{prior}}_{\omega_f\to\tilde{\omega}_f}} 
   \: \exp \big( - \Delta\mc F \big) \right\}.
\end{equation}

\paragraph*{Full saddle point approximation.}
In the full saddle point approximation, the potential $\omega_f$ is given
by $\omega_f^*$ in Eq.\ (\ref{eq:saddle_omega}), where $\phi_f^*$ is determined
by Eq.\ (\ref{eq:density}) with $\omega_f^* = i W_f^*$.  In order to solve
(\ref{eq:saddle_omega}) and (\ref{eq:density}) for $\tilde\omega_f$ we
implement a relaxation scheme based on the evolution equation 
\begin{equation}
\label{eq:ddfs}
\frac{d\omega_f}{dt} = \frac{\delta\mc F}{\delta\phi_f} =: F.
\end{equation}
In discretized form the resulting iterative scheme is given by 
\begin{equation}
\omega_f^{(n+1)}=\omega_f^{(n)} + \Delta t \; F^{(n)},
\end{equation}
where $n$ denotes the index of the iteration step, $\Delta t$ is the step
length, and $F$ represents the ``driving force'' for relaxation.  The step
length should not be set too large, otherwise the iteration does not converge.
Usually a few tens of iteration steps are required to reach the fixed point
corresponding to the saddle point $\omega_f^*$.  In practice, however, we do
not need to know the exact saddle point solution at all times due to the
statistical property of the system. Therefore, it is sufficient to perform one
or two iteration steps, using the potential obtained in the MC-last step as the
initial one.  This procedure corresponds to a dynamic density functional scheme
(DDFT), where the fields $\omega_f$ evolve according to the artificial dynamics
(\ref{eq:ddfs}). If the characteristic time scale of this dynamics is much
smaller than the characteristic time scale of the particle dynamics, the fields
$\omega_f$ adjust to the particle configurations almost adiabatically. The
fluctuations of the potential during the simulation reflect the fluctuation of
p-chains, and the statistical fluctuations of the field are fully neglected.

\subsubsection*{iii. Identity switches}

The identity switches (i.e.\ switches between p- and f-chains) are driven by
the spin variables on the chain. To change the spin configuration one can
randomly choose a bead and randomly update its spin variable, and then
determine the resulting species from the new spin configuration
\cite{Semigrand}.  As mentioned in subsection \ref{II_1}, however, a p-chain
corresponds to ($2^N-1$) possible spin configurations, and an f-chain only to
one. Consequently most of the attempted switches will convert a p-chain
into a p-chain again and this scheme becomes very inefficient for large $N$.
In order to increase the transition probability from p-chains to f-chains, we
therefore implement a spin-biased scheme: We first generate a random integer in
the interval $[0,n_p*M_\mathrm{mid}+n_f)$. If the integer is smaller than
$n_p*M_\mathrm{mid}$, a p-chain is picked, otherwise, an f-chain is chosen.  If
a p-chain is picked, we randomly choose a bead on one of the $n_p$ p-chains,
and then flip the spin variable of this bead with a Metropolis acceptance
probability. Here $M_\mathrm{mid}$ is an arbitrary integer, which we typically
choose equal to the number $N$ of beads in a chain, and this bias scheme
amounts to attempting $M_{\mbox{\scriptsize mid}}$ times more spin flips
on p-chains than on f-chains.

We first consider the case where a spin flip is attempted on an f-chain
$\alpha$. By definition, all spin variables on this chain are zero before the
flip. If one variable $\tau_{\alpha j}$ is flipped to $\tau_{\alpha j}=1$, the
f-chain is eliminated from the system so that the f-chain density becomes
$\tilde\phi_f(\mb r)=\frac{n_f-1}{n_f}\phi_f(\mb r)$ (the potential
$\omega_f(\mb r)$ is kept fixed), and a new p-chain configuration is generated
with bond vectors distributed according to a Gaussian with a variance given by
the squared bond length.  The priori probability for generating a given chain
configuration $\{ \mb R_\alpha^j \}$ is thus given by 
$P_{\mathrm{prior}} 
   \propto \exp\big(- \frac{n_t}{V} \mc F_g[\{ \mb R_\alpha^j \}]\big)$
with
\begin{equation}
\label{eq:Fg}
\mc F_g[\{ \mb R_\alpha^j \}] 
  = \frac{V}{n_t} \frac{N}{4} \sum_{\alpha j}
         (\mb R_\alpha^j - \mb R_\alpha^{j-1})^2
     - \frac{V}{n_t} \ln \mc Z_0. 
\end{equation}
The Metropolis acceptance rate for this switch must be chosen
\begin{eqnarray}
P_{fp} & = &
   \mathrm{min}\Bigg\{1, \exp\Big[\Delta\mu(\mb R_\alpha^j)
     -\sum_{l=0}^{N-1}\ln(e^{\Delta\mu(\mb R_\alpha^l)}+1)\nonumber\\
  && -\; \frac{n_t}{V}\big( \Delta\mc F - \mc F_g[\{ \mb R_\alpha^j \}] \big)
     +\ln M_{\mbox{\scriptsize mid}}\Big]\Bigg\},
\label{eq:Pfp}
\end{eqnarray}
where $\Delta\mc F=\mc F(n_p+1,n_f-1)-\mc F(n_p,n_f)$ is the free energy
difference associated with the identity switch. Note that by subtracting
$\mc F_g[\{ \mb R_\alpha^j \}]$ from $\Delta \mc F$, we have subtracted the
contribution of the chain connectivity both at the particle and field
level. They are already taken into account in the p-chain generating
process ($P_{\mathrm{prior}}$).

Next we consider the situation where a flip of a spin variable $\tau_{\alpha
j}$ is attempted on a p-chain. We must distinguish between three cases here.
First, we may have a transition from $\tau_{\alpha j}=0$ to $\tau_{\alpha
j}=1$. In that case, the $\alpha$-th chain always remains a p-chain, and the
associated acceptance probability is given by
\begin{equation}
P_{pp,0\to 1} = 
   \mathrm{min}\Big\{1, \exp\big[\Delta\mu(\mb R_\alpha^j)\big]\Big\}.
\end{equation}
Second, we may have a transition from $\tau_{\alpha j}=1$ to
$\tau_{\alpha j}=0$ with at least one other $\tau_{\alpha k}=1$
($k\neq j$). After switching, the $\alpha$-th chain would still be a
p-chain, and the acceptance probability is calculated as
\begin{equation}
P_{pp,1\to0} =
  \mathrm{min}\Big\{1, \exp\big[-\Delta\mu(\mb R_\alpha^j)\big]\Big\}.
\end{equation}
The third case corresponds to the transition from p-chain to f-chain.  This is
possible if all spin variables $\tau_{\alpha k}$ are zero for $k\neq j$ and the
spin variable $\tau_{\alpha j}$ switches from $\tau_{\alpha j}=1$ to
$\tau_{\alpha j}=0$. The $\alpha$th chain is then removed from the set of
p-chains and turned into an f-chain. Since f-chains do not have an explicit
configuration, its contribution to the field density is expressed as
$\phi_f(\mb r)/n_f$ resulting in the new density $\tilde\phi_f(\mb
r)=\frac{n_f+1}{n_f}\phi_f(\mb r)$. The potential $\omega_f(\mb r)$ is kept
fixed. The acceptance probability of this move is evaluated as
\begin{eqnarray}
P_{pf} & = &
    \mathrm{min}\Bigg\{1, \exp\Big[-\Delta\mu(\mb R_\alpha^j)
        +\sum_{l=0}^{N-1}\ln(e^{\Delta\mu(\mb R_\alpha^l)}+1)\nonumber\\
&& -\; \frac{n_t}{V}\big( \Delta\mc F + \mc F_g[\{ \mb R_\alpha^j \}]\big)
   -\ln M_{\mbox{\scriptsize mid}}\Big]\Bigg\},
\label{eq:Ppf}
\end{eqnarray}
where $\Delta\mc F=\mc F(n_p-1,n_f+1)-\mc F(n_p,n_f)$ is the free energy 
difference associated of the described identity switch, and the contribution
$\mc F_g$ (given by Eq.\ (\ref{eq:Fg})) of the chain connectivity has to
be subtracted from $\Delta \mc F$ just as in Eq.\ (\ref{eq:Pfp}) in order 
to satisfy the detailed balance condition. 

\section{Results and discussion}

We will now demonstrate and evaluate the algorithm at the example of a confined
homopolymer melt in slit geometry. Our model system has an accessible volume 
$V=L_x\cdot L_y\cdot L_z=8\cdot 8\cdot 16$ (in units of $R_g$) with
impenetrable hard walls at $z = \pm L_z/2$, and it contains $n_t=20000$ polymer
chains of length $N=20$ beads. The interactions between beads are repulsive
with excluded volume parameter $v_0=10$.  The total volume of the simulation
box is $L_x \cdot L_y \cdot 2 L_z$ and it is split into $n_x\cdot n_y\cdot
n_z=16\cdot 16\cdot 64$ cells. We use periodic boundary conditions in all
directions and implement the hard walls by applying a very large potential 
for $z$-values less than $-L_z/2$ or larger than $L_z/2$.

\begin{figure}[t]
\centerline{\includegraphics[angle=0,scale=0.6,draft=false]{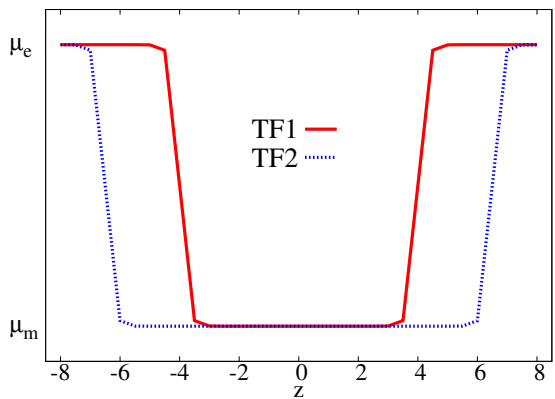}}
\caption{\label{fig:TF} Sketches of the TF obtained from Eq.~(\ref{TF1}) 
with $z_0 = 4$ (TF1) and $z_0 = 6.5$ (TF2)}
\end{figure}

The TF in the hybrid scheme is chosen as a tanh function with the form

\begin{equation}\label{TF1}
\Delta\mu(\mb r)=\frac{\mu_e+\mu_m}{2}
   +\frac{\mu_e-\mu_m}{2}\tanh\big[(|z| - z_0)/w\big]
\end{equation}

where $\mu_e$ and $\mu_m$ are the values of $\Delta\mu(\mb r)$ in the particle-
and field-dominated domains, respectively, $z_0$ controls the position of the
boundary between domains, and $w$ corresponds to the step width of
$\Delta\mu(\mb r)$ at the boundary. These parameters determine the profile of
the density distributions of p-chains and f-chains. In the following we set
$w=0.25$. The TF is homogeneous along the $x$ and $y$ directions, and the curve
along the $z$ direction is plotted in Fig.~\ref{fig:TF} for two choices of the
boundary position, $z_0 = 4$ (labelled TF1) and $z_0 = 6.5$ (labelled TF2).

In the simulations presented here, we typically performed 10000 Monte Carlo
steps (MC steps) to equilibrate the system and another 10000 MC steps to
evaluate the statistical averages. Each MC step comprises $M_p=n_p \times N$
particle configuration updates (i.e., one trial move per bead on average), $M_s
\sim 1000 - 20.000$ attempts to switch chain identities, and $M_f$ field
updates, where we chose $M_f = 1$ in the case of DDFT simulations, and $M_f
\sim 100-1.000$ if field fluctuations are included.  In some cases discussed
below, we also used $M_f=1/3$, meaning that the fields are updated only in
every third MC step. In each switch update, we switch one bead, such that
roughly $M_s/N*0.4 \sim 200$ chains are switched in one MC step for a typical
acceptance rate of 0.4. 

\begin{figure}[t]
\centerline{\includegraphics[angle=0,scale=0.5,draft=false]{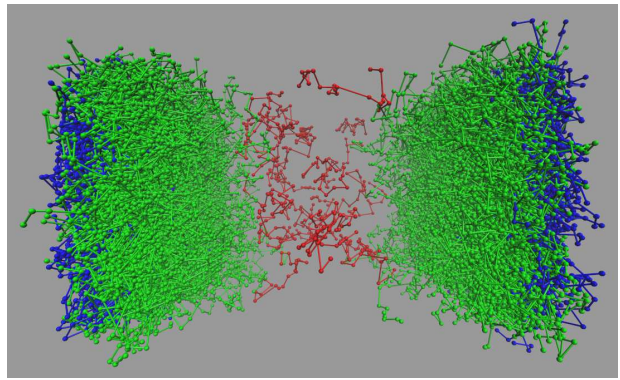}}
\caption{\label{fig:snapshot} Simulation snapshot for polymers resolved by 
particles obtained with the tuning function TF2 with $\mu_e=0$ and $\mu_m=-8$. 
Chains in the bulk region are colored red, chains close to the impenetrable
boundaries are colored blue, chains in the interfacial region between
particle and field domain are colored red. }
\end{figure}

Figure \ref{fig:snapshot} shows a typical simulation snapshot of the polymers
in particle representation in a system with tuning function TF2.  In the bulk
region, very few chains are resolved by particles. Close to the boundaries,
almost all chains are represented by particles. The polymers represented by
fields are not shown in this snapshot.

\subsection{Validation of the hybrid model}

\begin{figure}[t]
  \centering
  \subfigure[]{
    \label{figa:densityProfile} 
    \includegraphics[angle=0, width=8.0cm]{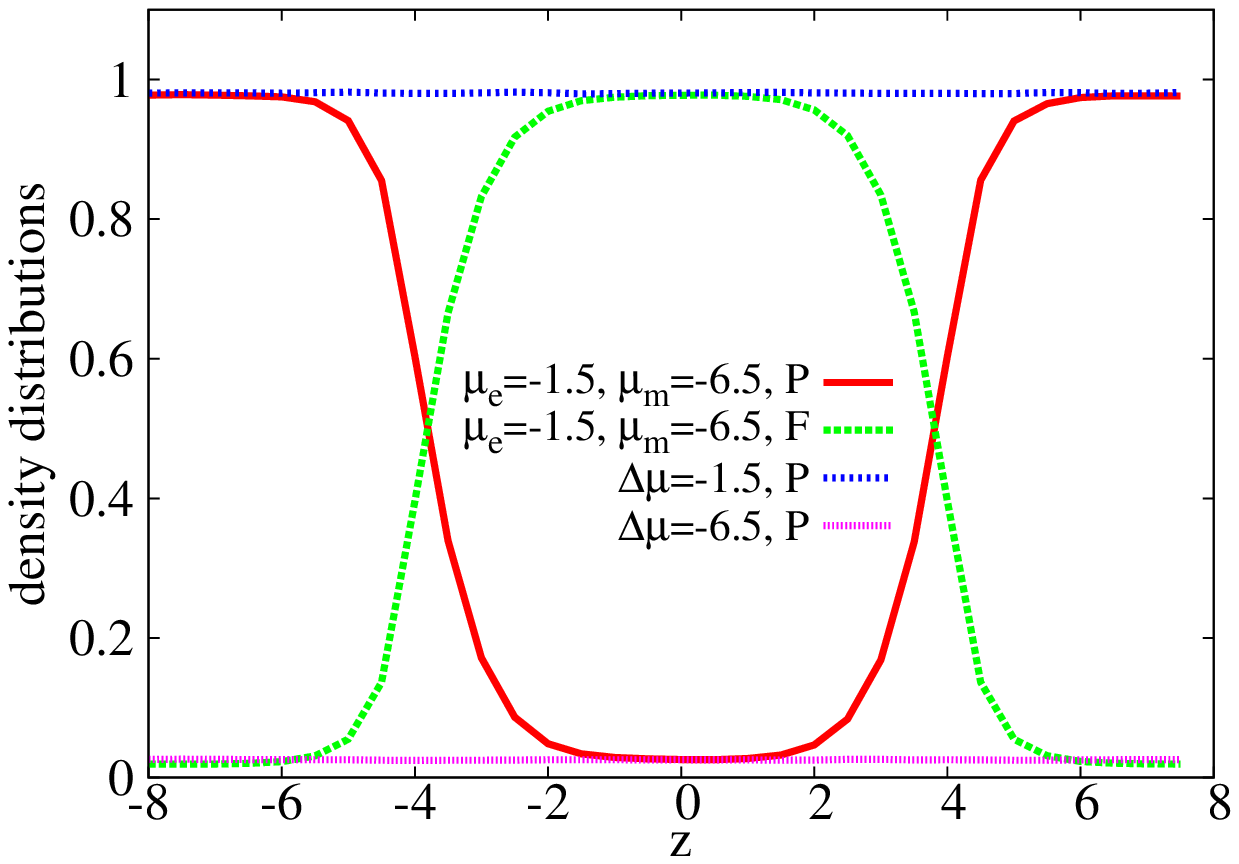}}
  \subfigure[]{
    \label{figb:densityProfile} 
    \includegraphics[angle=0, width=8.0cm]{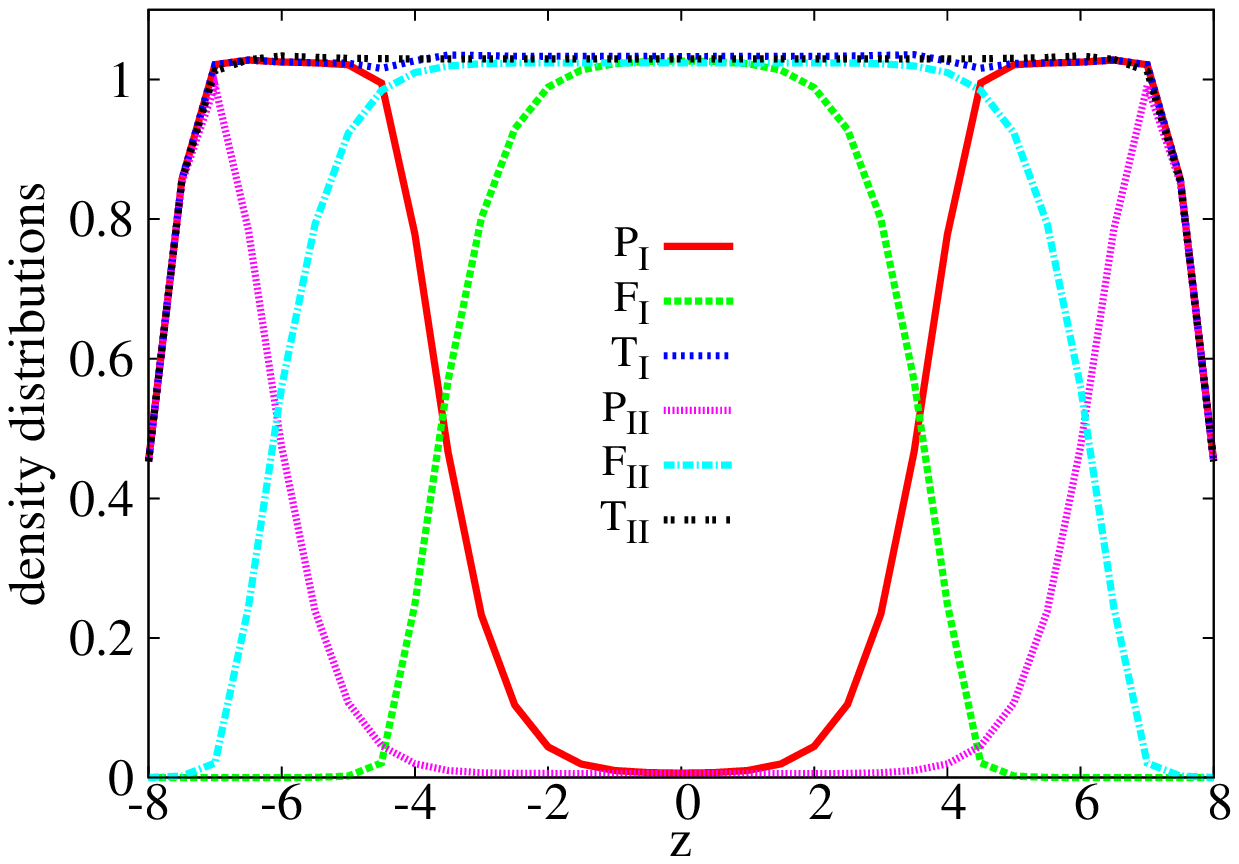}}
  \caption{Density distributions obtained with the hybrid model 
in a bulk system (a), and in a confined slit (b). In (a) the TF is chosen 
as TF1 with parameters $\mu_e$ and $\mu_m$ as indicated, and in (b)
the TFs are chosen as TF1 (I) and TF2 (II) with $\mu_e=0$ and $\mu_m=-8$.
Here ``P" refers to the density of p-chains, ``F" to the density 
of f-chains, and ``T" to the total density. }
  \label{fig:densityProfile} 
\end{figure}

We will first demonstrate that the chains can be successfully partitioned into
p-chains and f-chains, and that their densities are controlled by the TF. Fig.\
\ref{fig:densityProfile} compares the density distributions in the $z$
direction as obtained from hybrid simulations of the confined system (Fig.\
\ref{figb:densityProfile}) with the corresponding profiles in a bulk system of
size $L_x \cdot L_y \cdot L_z$ without confining walls (Fig.\
\ref{figa:densityProfile}).  In the bulk system, a homogeneous TF generates
homogeneous densities, and a large value of TF results in a large quantity of
p-chains and a small quantity of f-chains. Fig.~\ref{figa:densityProfile}
demonstrates that an inhomogeneous TF that switches between $\mu_e=-1.5$ and
$\mu_m=-6.5$ produces inhomogeneous densities of p-chains and f-chains, such
that the p-domain with $\Delta \mu = \mu_e$ is mostly occupied by p-chains, and
the f-domain with $\Delta \mu = \mu_m$ mostly by f-chains. In confined slit
geometry, sharp interfaces appear near the hard walls. By choosing the TF
suitably, one can enforce that polymers near the walls are represented by
particles, while in the middle of the system (bulk region), most of the chains
are represented by fields (Fig.\ \ref{figb:densityProfile}). For the tuning
function TF2, only a fraction of roughly 1/5 of all chains are p-chains (in the
case of TF1, it is about 1/2).  Fig.\ \ref{figb:densityProfile} also nicely
demonstrates the potential applications of the hybrid method. For regions where
a high resolution method is required, i.e., close to interfaces, one can use
the PR method for detailed investigations, while for regions where the
knowledge of density profile is sufficient, one can adopt the numerically
cheaper FR method. Since the high resolution PR region occupies only a small
part of the system, a hybrid simulation is more efficient than a pure particle
simulation. This will be discussed more quantitatively below.

The collective properties of p-chains and f-chains, representing the physical
properties of the whole system, are determined by the system parameters only
(i.e., the excluded volume parameter, the size of the box, and the total number
of grid points) and ideally, they should be independent of the concrete forms
of the TF. However, we have already discussed in Sec.\ \ref{II_1} that in
reality, the PR and FR representations are not fully equivalent for two
reasons: First, because of the mean-field approximation that has entered the
derivation of the hybrid model, and second, because of the finite grid size.
The grid size affects PR and FR simulations in different ways especially close
to interfaces and surfaces, where the density profiles vary strongly on the
scale of single grid cells. Since the hybrid model should reproduce the
properties of the particle model, the TF should be chosen such that such
interfacial regions are treated at the PR level.

\begin{figure}[t]
\centerline{\includegraphics[angle=0,scale=0.6,draft=false]{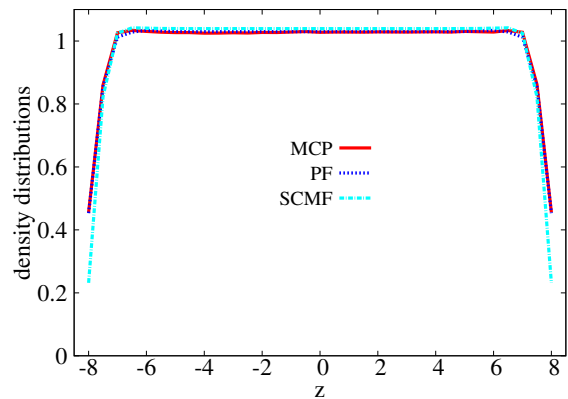}}
\caption{\label{fig:densityComparison} Total density distribution
in the confined system for the pure particle model (MCP), the pure field
theory (SCMF), and the hybrid model (PF) with tuning function TF2
(with $\mu_e=0$ and $\mu_m=-8.0$).}
\end{figure}

To test this, we focus on the density distributions and configurations of
polymers close to the surfaces of the slit. Due to the impenetrability
condition, the polymers are depleted near the wall.  Fig.\
\ref{fig:densityComparison} compares the total density distributions obtained
from the pure particle model (denoted MCP), the hybrid model (denoted PF), and
the pure field model, which in our case corresponds to a
standard self-consistent mean field theory (SCMF). As stated above, the pure
particle model represents the reference model, whose properties should be
reproduced by the other models. From Fig.\ref{fig:densityComparison}, it can be
seen that the density profiles in the pure field model are steeper close to the
boundaries than in the particle model, and the contact density is too low.
Note that the chain resolution and the space resolution are the same for MCP
and SCMF. The Edwards length $\xi_E$, which measures the range of fluctuation is an important
quantity in field-based simulations. Usually, the lattice spacing should be chosen smaller
than $\xi_E$. In the current work, $\xi_E$ is evaluated in units of $R_\mathrm{g}$ as
$\xi_E\sim 0.22R_g$ \cite{Fredrickson}, and it is a bit smaller than the lattice 
spacing $0.5R_\mathrm{g}$ but still on the same order.
We checked by varying the spacing size in the SCMF theory that the density curve is almost 
independent of the lattice spacing. Further it can be seen from Fig.\ref{fig:densityComparison} 
that the width of the depletion region near the boundary is about $R_\mathrm{g}$, 
two times larger than the lattice spacing, therefore the density difference
here is not a defect of finite lattice spacing.
This means that the pure field model cannot properly capture the properties of the
system close to the boundary. In contrast, the total density curve obtained
from the hybrid model is almost identical to that obtained from the pure
particle model. In the middle of the system, which is the bulk region, all
three density curves almost match. From another perspective, this means that
field-based simulations can be improved by treating selected regions in space
at the particle level within our hybrid scheme.

Next we inspect the polymer configurations, especially focussing on the
mean-squared radius of gyration (MSRG) of chains. This quantity is easily
accessible for p-chains, and it cannot be calculated for f-chains, therefore we
only consider p-chains here.  We cut the simulation box along the $z$ direction
into $n_z$ slices centered at $z_i$ for the $i$-th slice, and the MSRGs at
$z_i$ are defined as the corresponding averages of all the p-chains with their
centers in this slice. We note that there are other quantities that 
can be extracted from both p-chains and f-chains, such as the chain-shape function \cite{MC_field}.

\begin{figure}[t]
  \centering \subfigure[]{
    \label{fig:Rgza} 
    \includegraphics[angle=0, width=8cm]{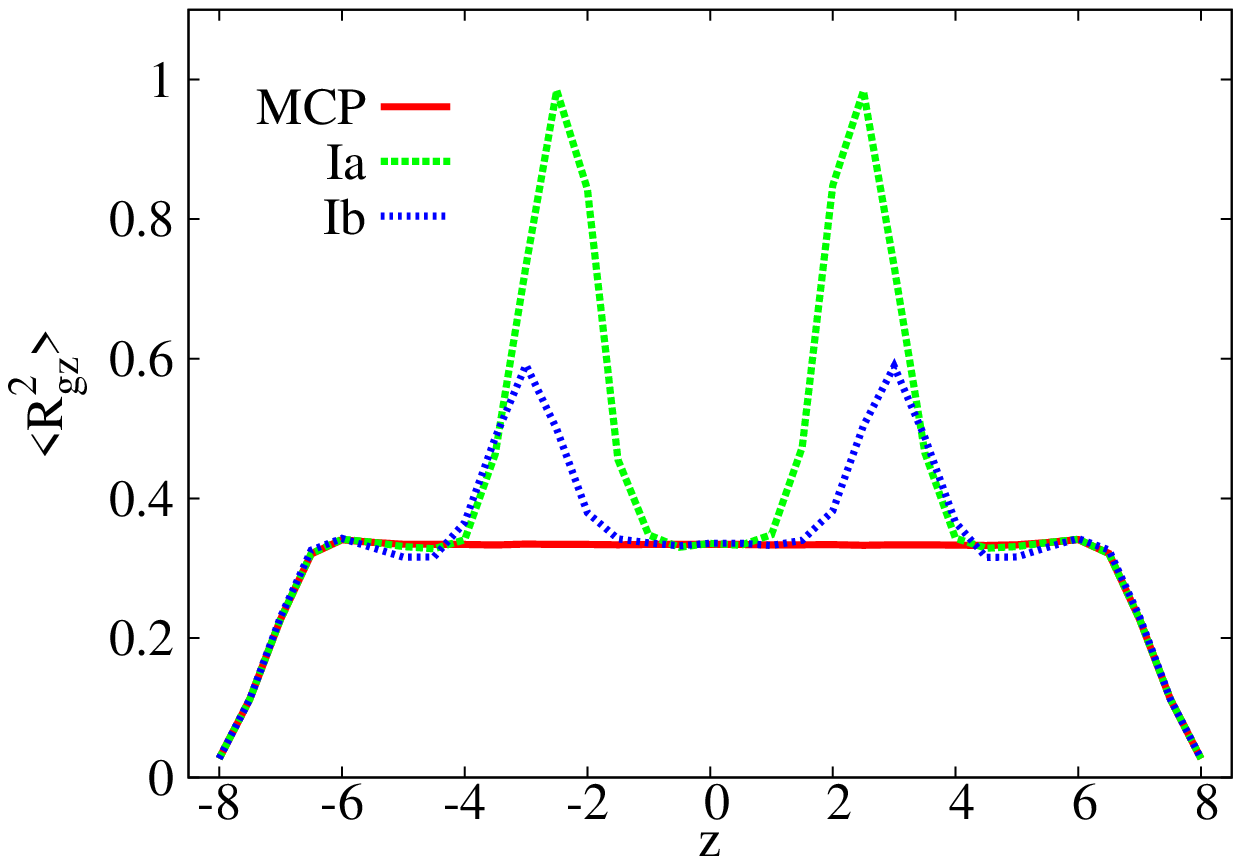}}
  \hspace{-0.5cm}
  \subfigure[]{
    \label{fig:Rgzb} 
    \includegraphics[angle=0, width=8cm]{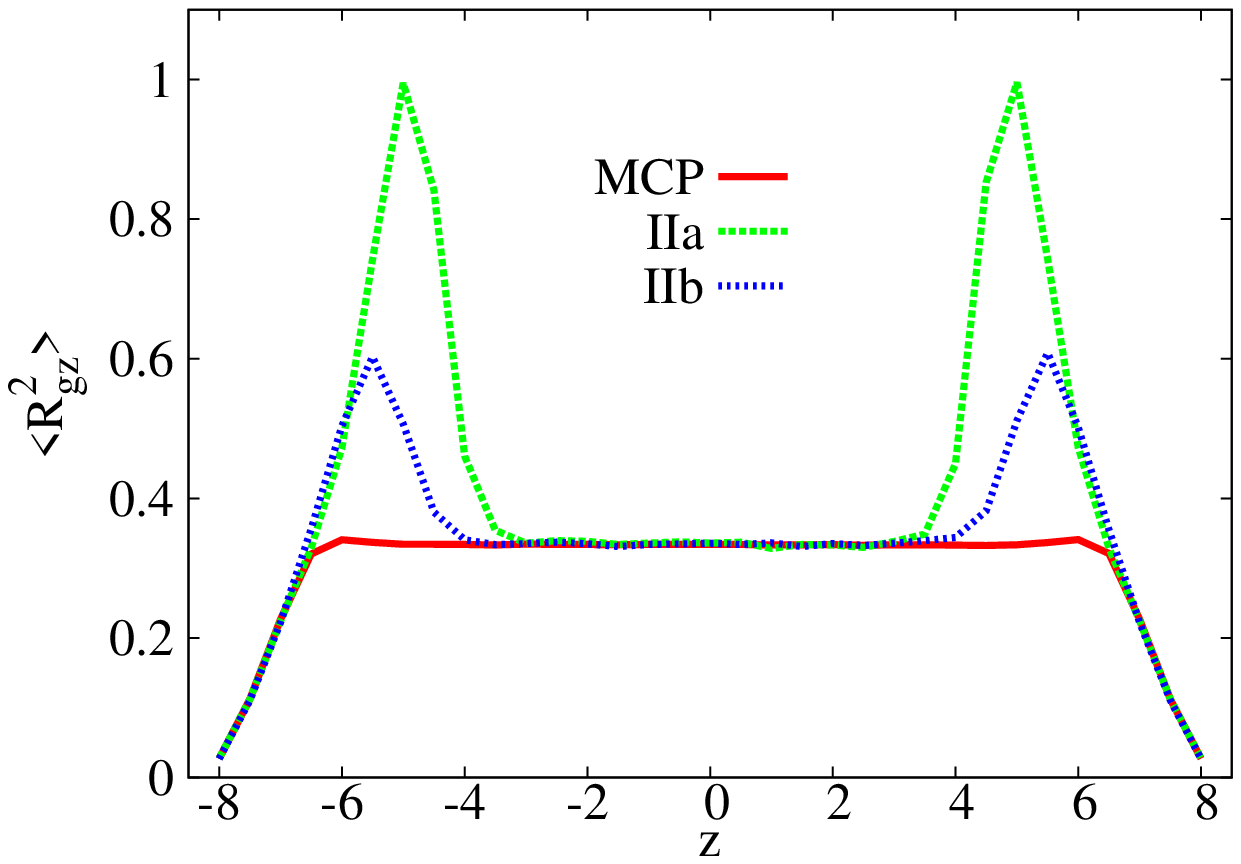}}
\caption{\label{fig:Rg} Profiles of the $z$-component of the MSRG of
p-chains in the pure particle model (MCP) and the hybrid model with 
tuning function TF1 (I) and TF2 (II). Here ``a" corresponds to 
$\mu_e=0$ and $\mu_m=-8$, while ``b" refers to $\mu_e=-3$ and $\mu_m=-7.0$.
Within a distance $\sim 2.5$ of the p/f-boundaries (at $z_0=\pm 6.5$),
the conformations of p-chains in the hybrid model are not representative
of the conformations of all chains.
}
\end{figure}

Figure \ref{fig:Rg} shows the MSRG profiles of p-chains along the $z$-direction
in the  pure particle model and the hybrid model with TF1 and TF2. Only the
profiles of the $z$-component of the MSRG ($\langle R_{gz}^2\rangle$) are
plotted -- the $x$- and $y$-components of the MSRG ($\langle
R_{gx,y}^2\rangle$) are constant throughout and assume the value $0.33$  which
is expected theoretically for Gaussian chains. The $z$-component reaches the
same value in the middle of the slab (bulk region), but close to the
boundaries, the polymers are a bit compressed.  This can be seen both in the
pure particle model and the hybrid model. The curves almost match close to the
wall -- in the p-domain of the hybrid model -- and in the middle of the slab --
in the f-domain.  In the intermediate region connecting p- and f-domains, the
MSRG of the p-chains in the hybrid model deviates from the MSRG measured in the
particle model: It is slightly reduced at the p-side of the p/f domain
boundary, and strongly enhanced at the f-side. This discrepancy can be
explained by the fact that the partitioning of chains into p- and f-chains
depends on their conformation in the p/f boundary region. On the f-side,
p-chains that stretch into the p-domain turn into f-chains with smaller
probability than p-chains that withdraw from the p-domain. Hence the remaining
p-chains are on average elongated in the $z$-direction. On the p-side, the
situation is the other way round: Chains that stretch into the f-domain turn
into f-chains with higher probability, and the remaining p-chains are slightly
compressed on average.  The effect can be reduced by adopting a smoother TF,
e.g., by reducing $\mu_e-\mu_m$, see the curves Ib and IIb in Fig.\ref{fig:Rg},
or by reducing $w$ (data not shown), but it never fully disappears.  Hence it
is important to realize that the p-chains in the p/f-boundary regions (within
2.5 $R_g$ of the boundary defined by the TF) are not representative of all
chains in the system and should not be used to determine conformational
properties of chains. This is a feature of the model, not an artefact.

More seriously, the total density also exhibits a very small dip in the
p/f-boundary region\cite{Hybrid_PF}.  Deviations of the total density are also
observed in other hybrid methods, and an additional potential is sometimes
introduced in order to remove these artefacts \cite{CandQ,PinningPotential}.
This is also an option in our model. In the present application to confined
homopolymer slabs, the effect is so small that an adjustment was not necessary.

\subsection{Numerical efficiency of the hybrid model}

Increasing computational efficiency is always the primary motivation for
introducing multiscale schemes, and we would like to show in the following that
the present hybrid method is more efficient than pure particle simulations.
The computational costs of a simulation based on a particle model
scales with the number of particles, while in a field description, it scales
with the number $M$ of grid points that are used to discretize space (or with
$M \ln M$ if Fourier methods are used).  Therefore, the hybrid model is most
efficient in dense systems, where the number of segments (beads) per grid cell
is large. At fixed number of grid points, one can choose a TF such that the
total number of p-chains is small, and this makes the simulation efficient. 

\begin{figure}[h]
  \centering \subfigure[]{
    \label{fig:timea} 
    \includegraphics[angle=0, width=8cm]{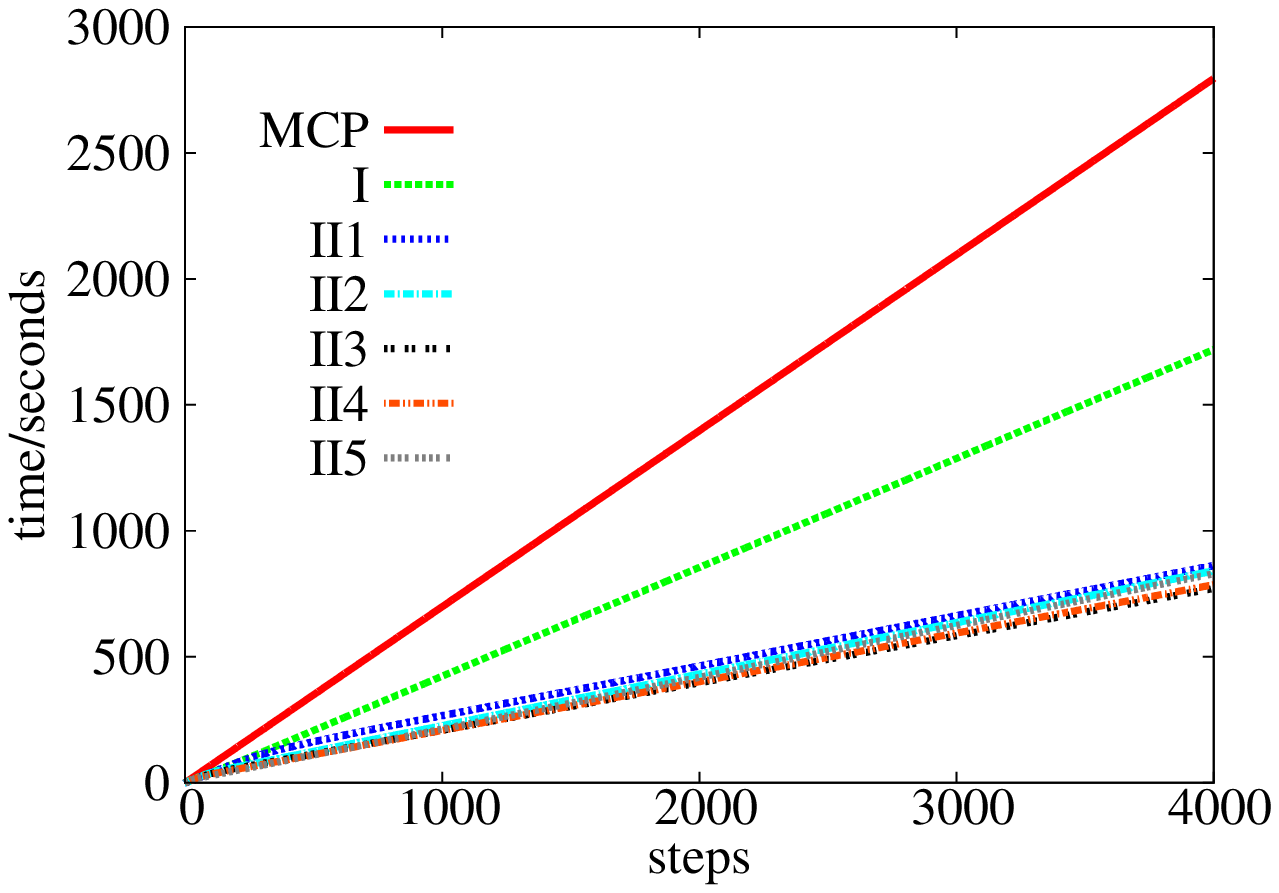}}
  \hspace{-0.5cm}
  \subfigure[]{
    \label{fig:timeb} 
    \includegraphics[angle=0, width=8cm]{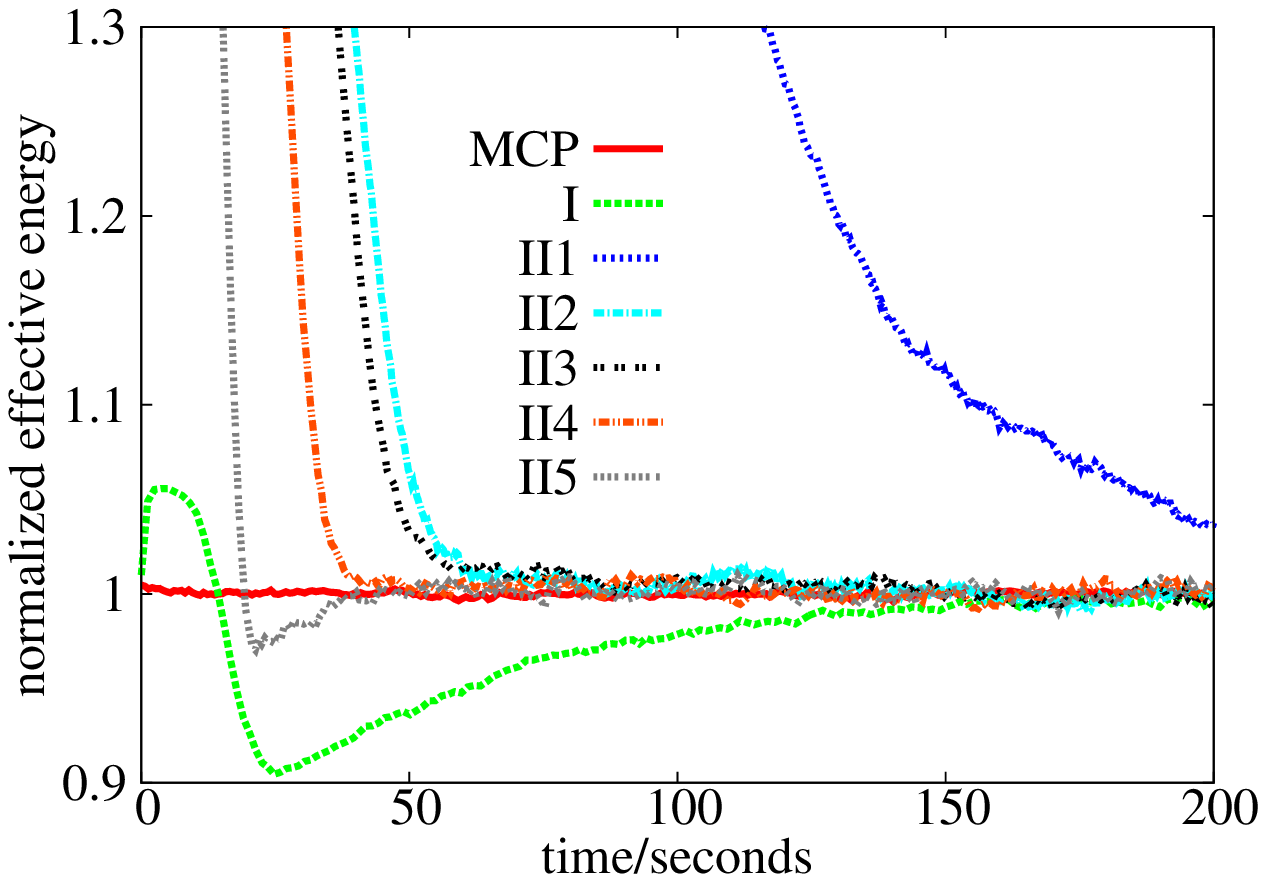}}
  \caption{(a) Computational costs per MC step and (b) normalized 
effective energy relaxation vs. CPU time for pure particle simulations (MCP),
and hybrid simulations with tuning function TF1 (I) and TF2 (II) using
the parameters $\mu_e=0$, $\mu_m=-8$. The field step length is $dt=0.1$. 
The parameters are: $M_f=1/3$, $M_s=20000$ (I); $M_f=1$, $M_s=2500$ (II1), 
$M_f=1$, $M_s=7000$ (II2), $M_f=1/3$, $M_s=7000$ (II3), 
$M_f=1/3$, $M_s=10000$ (II4), and $M_f=1/3$, $M_s=18000$ (II5). 
}
  \label{fig:time} 
\end{figure}

In order to evaluate the efficiency of our hybrid scheme, we use the same
methods to update the particle configurations in the pure particle simulation
and the hybrid simulation, and perform the calculations on the same desktop
computer (i7 CPU, 2.67 GHZ). We find that in each MC step, the
computational cost for updating the chain configurations dominates the total
costs, hence the choice of $M_f$ (i.e., whether we choose $M_f=1$ or $M_f =
1/3$) and $M_s$ has little influence on the overall computational costs.
However, the choice of $M_s$ does strongly affect the equilibration time for
the effective energy.  From Fig.~\ref{fig:timea}, one can see that the hybrid
calculation with TF1, where about half of the chains are p-chains, is roughly
1.4 times faster than the pure particle simulation, while one can achieve a
speedup of a factor 3.5 with TF2, where about 1/5 of all chains are p-chains.
Reducing $M_f$ by a factor of 3 has only a small influence on the computational
costs.

Figure~\ref{fig:timeb} shows the relaxation of the effective free energy to the
equilibrium value. The relaxation is almost independent of $dt$ in the present
system, hence we choose $dt=0.1$ for all cases. It can been seen that the pure
particle method converges very rapidly, while the relaxation in the hybrid
simulation is a bit slower, since additional time is necessary to equilibrate
the partitioning between p-chains and f-chains. Although it takes more time to
equilibrate the system within the hybrid method, this equilibration period is
still very short (about 100 CPU seconds) compared to the much longer times
(about 6000 CPU seconds) required for calculating the statistical averages. The
hybrid method develops its high efficiency in this second stage, where
data are accumulated.  If the computational costs would strictly
scale with the number of particles, we would reach a theoretical speedup of a
factor of 5 in our system. This number is reduced due to the overhead costs for
bookkeeping, field updates and switch moves (separating these costs is
difficult), but we could still reach a speedup factor of 3.5. In systems where
the fraction of particle resolved domains is smaller, the speedup will be even
larger.

When judging the efficiency of a computational algorithm, another aspect that
has to be considered is the complexity of the technique, and the man-hours
needed to implement it. The ultimate goal is always to devise simple algorithms
with high computational efficiency, although this is usually hard to achieve in
practice. The new constituent in the present scheme compared to the usual
particle-based MC scheme and field-based theory (e.g., self-consistent field
theory) is the particle switching. Thus the hybrid scheme can be implemented by
adding particle-switching algorithms that are executed after updating particle
positions and fields. The switching algorithm is short and not complex, and the
man-hours necessary for implementing it are less than or comparable to those
required for writing a code for its pure particle-based counterpart. From a
practical point of view, the main challenge is to combine particle and field
simulation methods in one single package -- i.e., introduce FR simulation
methods in codes designed for particle simulations and vice versa. However, as
explained in the introduction, there is generally a growing interest in hybrid
schemes that combine particle and field methods. Within a package that can
already combine particle and field-based (DDFT) simulation schemes,
the implementation of our adaptive hybrid scheme should not require much
effort.

\subsection{Incorporation of fluctuations and outlook}

\begin{figure}[h]
\centerline{\includegraphics[angle=0,scale=0.6,draft=false]{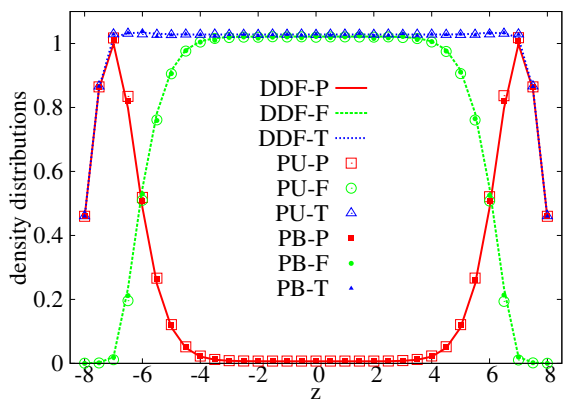}}
\caption{\label{fig:PU_PB} Densities of p-chains (P), f-chains (F) and the 
total density (T) obtained with the dynamic density functional scheme (DDFT), 
the potential unbiased algorithm (PU), and the potential biased algorithm (PB). 
The tuning function is chosen as TF2 with $\mu_e=0$ and $\mu_m=-8.0$. 
The noise amplitudes in the PU and PB schemes are chosen such that the
acceptance probability is about 0.3. The densities obtained with these
three schemes almost match each other.
}
\end{figure}

In all of the above calculations, only the fluctuations of p-chains were taken
into account, and those of the fields were neglected. Fluctuations sometimes
play an important role in determining the properties of a dilute solution or a
dense system near the critical point of a phase transition. In general one
should devise methods that account for fluctuations whenever necessary. The
present hybrid method does not abandon all fluctuation effects in principle,
rather they are treated in different ways due to the difference of types of
degrees of freedom. The configurations are sampled by the particle based method
through particle moves, while the fields are sampled by the field based method
through the evolution of fields. We have up to now focused on methods that
neglect fluctuations in the FR part. In the following, we incorporate the
fluctuations to the field by using the unbiased and biased potential sampling
schemes (see subsection \ref{Simulation}). We choose a small size of box in
the $x$ and $y$ directions with also a small number of grid points, i.e.,
$L_x=L_y=4$, $n_x=n_y=8$. The number of steps in updating the fields in each MC
step is set $M_f=100$. Figure \ref{fig:PU_PB} shows the comparison of the
density distributions obtained from the DDFT scheme, the PU and PB schemes. The
parameters are chosen as $\mu_e=0$, $\mu_m=-8$. The results are independent of
the step length $dt$. The curves obtained from these three different schemes
are almost identical, which is expected, since the density of the polymers in
the system is very high, and the fluctuation effect is expected to be small.
In our applications, we find that the PB schemes is not as efficient as the PU
and DDFT schemes. This may be different in more complex systems, where the bias
of PB moves towards following a free energy gradient may be of advantage.

Even if we use advanced Monte Carlo techniques such as the PB scheme to sample
the field fluctuations, they do not represent the true dynamics of the system.
In order to study rheological and dynamical properties, the hybrid scheme
should be complemented with a consistent dynamical model. In particular, mass
(particle number) and momentum should be locally conserved. Particle models
with ``realistic" dynamics satisfy this requirement. For example, overdamped
Brownian particle simulations conserve the mass, while dissipative particle
dynamics conserves both mass and momentum. On the side of the field model, the
Maurits-Fraaije \cite{density_dynamics} density dynamics conserves mass, at
least approximately, and such DDFT models can be extended such that momentum is
conserved as well \cite{zhang_11}. Difficulties in a true hybrid dynamics model
arise from the requirement of mass and momentum conservation during the process
of resolution switching. Several works have tried to address these notoriously
difficult issues in hybrid models, see, e.g., Refs.\
\cite{particle_FH,MD_FHM,MD_SDPD,MD_MCD}.  It is conceivable that particle-based
Brownian simulations, combined with field-based density dynamics and a
particle-field switching process which is compatible with the continuity
equation for the mass, could capture the dynamics of the system at the same
level as pure particle-based Brownian simulations.  The construction of such
a model is currently under way.

\section{Summary and remarks}

In the present paper, a hybrid particle-continuum simulation scheme with
adaptive resolution for soft matter systems was derived based on a
field-theoretical approach. In such a hybrid resolution method, polymer chains
can switch their representations on the fly according to a pre-determined
tuning function TF. A proper form of TF can ensure that polymer chains are
described by particles in small selected regions in order to directly observe
the configuration dependent properties, and by density profiles in the
remaining large bulk regions, where configuration details are not interesting.
Therefore, the hybrid scheme is computationally more efficient than the pure
particle method, while keeping the physical accuracy in the particle regions.
The hybrid approach is particularly attractive for simulations of dense systems
where particle simulations become expensive compared to field-based
simulations.

In the present formulation of the hybrid model, the high-resolution domains
have to be determined beforehand. They may follow polymer-solid interfaces,
e.g., in applications to polymer nanocomposites, but they do not explicitly
depend on the polymer configuration. Hence our scheme is adaptive, but not
self-adaptive. Constructing a self-adaptive scheme is another major challenge
for the future.  Physically, self-adaptive means that the TF should be
determined by the system itself. So far, the shape of the TF is essentially a
tanh function with pre-defined amplitude, offset, interface position and step
width.  None of these parameters is determined by the system itself.  However,
the optimal TF depends on  the intrinsic properties of the hybrid system. For
example, due to the asymmetric property of the p-chains and f-chains, a
non-zero reference value of $\Delta \mu$ where the fraction of p- and f-chains
is equal can be calculated which can then be used as offset. Concerning the
step width and the locations of the p/f interfaces, one possible and practical
way is to directly relate them to the local density and/or the local density
gradient.  The tuning function could be chosen such that it has large values in
interfacial regions (where the total density curve has a large slope) and low
values in the bulk (where the slope is zero). This should give the desired
result that p-chains aggregate in the interface while f-chains are dominant in
the bulk. Such hybrid models could then be called self-adaptive resolution
models in the sense that the chains described by different resolution models
can switch and redistribute according to the local density. We will explore
this option in future work.

The present hybrid model is constructed such that it targets polymer solutions.
However, the basic idea of the hybrid model can be applied to other systems.
The approach is especially suitable for very large systems with large bulk
regions and small interface regions or small regions that need detailed
investigation. One example is a polymer brush system, in which a small number
of brush polymers are attached to a plate, while a large number of polymers
remain in the bulk.  For such a system, the hybrid model is recommended. The
brush polymers can be treated by higher resolution method, i.e., the PR method,
to catch the more detailed properties (for example, the lengths of chains),
while the large bulk region can be treated by lower resolution method, i.e.,
the FR method, to save time.

The studies of a homopolymer solution using the hybrid scheme show good
agreement with those by the reference pure particle method, and the hybrid
scheme has demonstrated its higher computational efficiency. However, so far,
it can only be used to study static properties. In future work, we will extend
this hybrid scheme such that it can also be used to study dynamical properties
and flow behavior in complex fluids.

\bigskip
\begin{center}
\textbf{ACKNOWLEDGMENTS}
\end{center}

We thank Stefan Dolezel, Sebastian Meinhardt, Liangshun Zhang, and Jiajia Zhou
for helpful discussions and suggestions. This project is supported by the
German Science Foundation (DFG) within project C1 in SFB TRR 146. Simulations
were run on the computer cluster Mogon at the University of Mainz.

\end{document}